\documentclass[%
 reprint,
 amsmath,amssymb,
 aps,
]{revtex4-1}
\usepackage{float}

\usepackage{verbatim} 
\usepackage{graphicx}
\usepackage{dcolumn}
\usepackage{bm}
\usepackage{hyperref}
\usepackage{amsmath}
\usepackage{mathtools}

\usepackage{xcolor}
\hypersetup{
    colorlinks,
    linkcolor={red!50!black},
    citecolor={blue!50!black},
    urlcolor={blue!80!black}
}

\usepackage{color}
\usepackage{tikz}
\usetikzlibrary{shapes}

\begin{document}

\newcommand{\markerone}{\raisebox{0.5pt}{\tikz{\node[draw,scale=0.4,regular polygon, regular polygon sides=4, magenta,fill=none](){};}}}

\newcommand{\markertwo}{\raisebox{0.5pt}{\tikz{\node[draw,scale=0.4,regular polygon, regular polygon sides=4, green,fill=green](){};}}}

\newcommand{\markerthree}{\raisebox{0.5pt}{\tikz{\node[draw,scale=0.4,circle,blue, fill=none](){};}}}

\newcommand{\markerfour}{\raisebox{0.5pt}{\tikz{\node[draw,scale=0.4,circle, red,fill=red](){};}}}

\newcommand{\markerfive}{\raisebox{0.5pt}{\tikz{\node[draw,scale=0.3,regular polygon, regular polygon sides=3,cyan,fill=none,rotate=0](){};}}}


\title{Stick-Slip Dynamics of Migrating Cells on Viscoelastic Substrates}

\author{Partho Sakha De}
\author{Rumi De}%
 \email{Corresponding author: rumi.de@iiserkol.ac.in}
\affiliation{%
 Department of Physical Sciences, Indian Institute of Science Education and Research Kolkata, Mohanpur 741246, West Bengal, India\\}%


\begin{abstract}
Stick-slip motion, a common phenomenon observed during crawling of cells, is found to be strongly sensitive to the substrate stiffness. Stick-slip behaviours have previously been investigated typically using purely elastic substrates. For a more realistic understanding of this phenomenon, we propose a theoretical model to study the dynamics on a  viscoelastic substrate. Our model based on a reaction-diffusion framework, incorporates known important interactions such as retrograde flow of actin, myosin contractility, force dependent assembly and disassembly of focal adhesions coupled with cell-substrate interaction. We show that consideration of a viscoelastic substrate not only captures the usually observed stick-slip jumps, but also predicts the existence of an optimal substrate viscosity corresponding to  maximum traction force and minimum retrograde flow which was hitherto unexplored. Moreover, our theory predicts the time evolution of   individual bond force that characterizes the stick-slip patterns on soft versus stiff substrates. Our analysis also elucidates how the duration of the stick-slip cycles are affected by various cellular parameters.
\end{abstract}

\maketitle


\section{\label{sec:level1}Introduction}

Cell motility plays a key role in many important biological processes such as wound healing, morphogenesis, embryonic development, tissue regeneration to name a few \cite{Gardel_Rev_2010, Ridley_Sci_2003, Friedl_Nat_Rev_2009, Dahlin_1988, Walter_1971}. Though motility is expressed in multiple ways, crawling happens to be the most common form of movement for eukaryotic cells. During crawling, cell forms  protrusions at the leading edge which pushes the membrane forward and as a consequence the membrane exerts a backward force on the polymerising actin filaments, resulting in them \lq slipping' rearward towards the cell center, in a process known as  retrograde flow \cite{Ananta_2007,Lin_1995,Jurado_2005}. This process is accompanied by growth and strengthening of focal adhesions between the cell and the substrate which slow down the actin retrograde flow \cite{Greenberg_ostap_2016}. Thus, it allows actin polymerization to advance at the leading edge and in turn, the rate of translocation of the cell increases \cite{Gardel_Rev_2010, Jurado_2005, Mitchison_1988, Hu_2007}. The dynamic variation in retrograde flow coordinated  with assembly and disassembly of focal adhesions lead to stick-slip motion. 

Stick-slip is a kind of jerky motion that  has been found not only in living systems but also in passive systems such as peeling of scotch tapes \cite{Rumi_PRE_2004, Rumi_PRE_2005, Rumi_lecture}, earthquakes \cite{Burridge_earthq_1967, Rumi_earthq, Rumi_lecture} to name a few. Stick-slip behaviour is characterised by the system spending most of it's time in the \lq stuck' state and comparatively a short time in the \lq slip' state. During crawling of cells, stick-slip dynamics  has been experimentally observed on multiple occasions. Experiments on migrating epithelial cells showed that in the lamellipodium region the traction force decreases with increasing velocity inferring a stick-slip regime of the actin-adhesion interaction \cite{Gardel_retrograde_2008}. Stick slip motion has also been observed in embryonic chick forebrain neurons \cite{Chan_Science_2008}, migrating human glioma cells \cite{Bangasser_natcom_2017} and also in human osteosarcoma cells \cite{Aratyn_2010}. In case of fish  keratocytes, stick-slip kind of behaviours have been found in modulation of the cell shape  during crawling \cite{Lacayo_2007, Lee_nature_1993, Barnhart_biophys_2010}. Moreover, recent experiments show that the stick-slip dynamics is strongly affected by altering the substrate stiffness. For example, a close inspection of the leading-edge motion of crawling and spreading mouse embryonic fibroblasts revealed that the periodic lamellipodial contractions are vastly substrate-dependent \cite{Giannone_2004}. Further, in recent studies, cell motility found to be maximum and actin flow rate minimum at an optimal substrate stiffness \cite{Chan_Science_2008, Stroka_2009, Peyton_2005, Bangasser_natcom_2017, Alberto_2018}.

There are many theoretical studies that have contributed significantly to understand the cell migration process \cite{Rangarajan_2008, Dimilla_1991, Shemesh_2012, Danuser_2013, Mogilner_2015, Kabaso_2011, Kruse_Joanny_Prost_2006, Rubinstein_Mogilner_2009}. Quite a few models have also been proposed to unravel the stick-slip mechanisms. Such as, Barnhart {\it et. al.} deveoped a mechanical model to predict periodic shape change during keratocyte migration caused by alternating stick-slip motion at opposite sides of the cell trailing edge \cite{Barnhart_biophys_2010}. Also, leading edge dynamics, spatial distribution of actin flow, and demarkation of lamellipodium-lamellum boundary have been studied \cite{Mogilner_2015, Shemesh_2012}. Besides, simple stochastic models have provided a great deal of information on cell crawling. Stochastic bond dynamics integrated with traction stress dependent retrograde actin flow could capture the biphasic stick-slip force velocity relation \cite{Li_Dinner_2010, Sabass_2010}. There are other models based on stochastic motor-clutch mechanisms which have  provided many insights into substrate stiffness dependent migration process \cite{Chan_Science_2008, Bangasser_biophys_2013, Bangasser_natcom_2017, Alberto_natmat_2014, Alberto_2018}. As observed in experiments, these studies reveal the existence of an optimal substrate stiffness which found to be sensitive to cell motor-clutch parameters. However, most of these studies on rigidity sensing so far are either focused on purely elastic substrate \cite{Bangasser_biophys_2013, Bangasser_natcom_2017} or on purely viscous substrate \cite{Bennett_PNAS_2018}; whereas, physiological extracellular matrix is viscoelastic in nature. Also, recent experimental studies further reveal that the dynamics is greatly sensitive to the substrate viscosity \cite{Bennett_PNAS_2018, Chaudhuri_2015, Lautscham_2014}.

In this paper, we present a theoretical model of the leading edge dynamics of crawling cells on a viscoelastic substrate. Our theory based on a framework of reaction-diffusion equations takes into account the retrogate flow of actin, myosin contractility, force dependent assembly and disassembly of focal adhesions integrated  with cell-substrate interaction. The model predicts how these cellular components work together to give rise to the spontaneous emergence of stick-slip jumps as observed in experiments. More importantly, it elucidates the effect of variation of substrate viscoelasticity on the \lq stick-slip' dynamics. Interestingly, it predicts, the existence of an optimal substrate viscosity corresponding to  maximum traction force and minimum retrograde flow as observed in case of elastic substrate  \cite{Chan_Science_2008, Bangasser_biophys_2013, Bangasser_natcom_2017}.  Moreover, our continuum model framework captures the time evolution of individual bond force that has remained unexplored so far. These findings suggest that the nature of non-trivial force loading rate of individual bonds play a crucial role in determining the stick-slip jumps and thus explain the distinctive patterns on varying substrate compliance. Our theory also provides an analytical understanding of how the cellular parameters such as substrate stiffness, myosin activity, retrograde flow affect the  duration of the stick-slip cycles.

\section{\label{sec:level1}Theoretical Model}

\begin{figure}[!t]
\begin{center}
\includegraphics[scale=0.45]{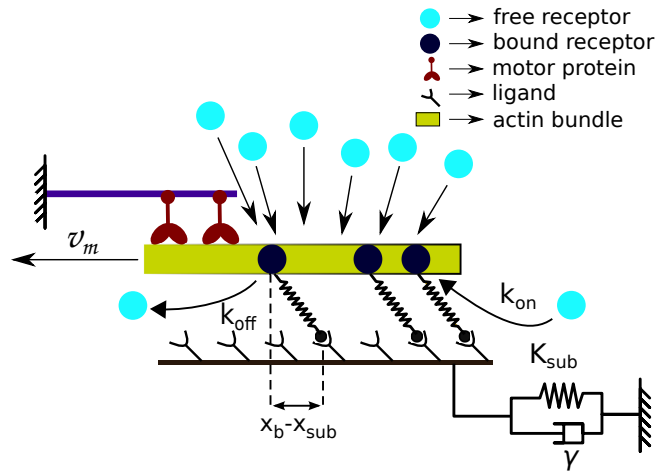}
\caption{Schematic diagram of the model: free receptors (denoted by light circles) diffuse within the actin cytoplasm.  The free receptors bind with  ligands on the substrate to form the bound receptors (dark circles), that forms the ligand-receptor bonds. The F-actin filament is pulled by myosin motors (maroon structures) with the retrograde flow velocity, $v_{m}$.  Viscoelastic substrate is modelled by a spring and a dashpot. (color online)}
\label{model_illustration}
\par\end{center}
\end{figure}

Our theory is based on the molecular clutch mechanisms proposed to describe the transmission of force from actin  cytokeleton to extra cellular matrix \cite{Mitchison_1988, Chan_Science_2008, Alberto_natmat_2014}. The clutch module or the \lq connector' proteins provide a dynamic link between F-actin and adhesion complexes and slow down the F-actin retrograde flow \cite{Gardel_Rev_2010, Ananta_2007, Jurado_2005}. Our model consists of free receptors representing these \lq connector' proteins diffusing in the actin  cytoplasm. The substrate consists of a large number of adhesive ligands which can bind with these free receptors  to form closed bonds as illustrated in Fig. \ref{model_illustration}. Thus, the receptors are considered to be either in free or bound states denoting  open or closed ligand-receptor adhesion bonds. The ligand-receptor bonds are modelled as Hookean springs of stiffness $K_{\rm c}$. As the F-actin bundle (modelled as a rigid rod), pulled by the myosin motors \cite{Greenberg_ostap_2016}, moves with the retrograde velocity, $v_{\rm m}$, the spring gets stretched and thus, the force on a single bond is given by multiplying the spring stiffness with the bond elongation as, $f=K_{\rm c}\left(x_{\rm b}-x_{\rm sub}\right)$; where $x_{\rm b}$ is the displacement of one end of the bond attached to the actin bundle and  $x_{\rm sub}$ is the displacement of the substrate (where the other end of the bond is attached). The retrograde flow velocity of the F-actin bundle slows down with increase in the force on the closed bonds and is given by the relation, $v_{\rm m}=v_{\rm 0}\left(1-\frac{F_{\rm b}^{\rm total}}{F_{\rm stall}}\right)$; where $v_{\rm 0}$ is the unloaded velocity,  $F_{\rm b}^{\rm total}$ is the total traction force due to all closed bonds, and $F_{\rm stall}$ is the total force exerted by myosin motors.
Here, $F_{\rm stall} = n_{\rm m} * F_{\rm m}$, where  $n_{\rm m}$ is the number of myosin motors present and $F_{\rm m}$ is the force exerted by an individual myosin motor \cite{Alberto_natmat_2014, Chan_Science_2008, Gardel_retrograde_2008, Greenberg_ostap_2016}.

We express the reaction between the free receptors and the ligands to form the bound receptors as
\begin{center}
\footnotesize 
\begin{eqnarray*}
R_{\rm f}\quad + \quad L \quad & \xrightleftharpoons[k_{\rm off}(f)]{k_{\rm on}(f)} & R_{\rm c}\\
{}[\rho{}_{\rm f}]\qquad \quad [\rho_{\rm L}]&  &  [\rho_{\rm c}]
\end{eqnarray*}
{\footnotesize \par}
\end{center}

where $\rho_{\rm f}$, $\rho_{\rm c}$, and $\rho_{\rm L}$ denote the densities  of free receptors, bound receptors, and ligands on the substrate respectively.  
(The total number of free and bound receptors is taken to be constant, {\it i.e.}, 
$\int_{0}^{\rm L}\left(\rho_{\rm c}+\rho_{\rm f}\right)dx=N_{\rm t}$ (constant). 
Now, the time evolution of the density of the free and the bound receptors are described by the following coupled reaction-diffusion equations, 

\begin{eqnarray}
\frac{\partial\rho_{\rm f}}{\partial t} & = & D\frac{\partial^{2}\rho_{\rm f}}{\partial x^{2}}-k_{\rm on}(f)\rho_{\rm f}+k_{\rm off}(f)\rho_{\rm c} \label{free} \\
\frac{\partial\rho_{\rm c}}{\partial t} & = & v_{\rm m}\frac{\partial\rho_{\rm c}}{\partial x}+k_{\rm on}(f)\rho_{\rm f}-k_{\rm off}(f)\rho_{\rm c}. \label{closed}
\end{eqnarray}

Here, the first term on the R.H.S of Eq. \ref{free} represents the diffusion of free receptors in the actin cytoplasm. The last two terms are the reaction terms of formation of  bound receptors and free receptors with respective reaction rates. For Eq. \ref{closed}, the first R.H.S term denotes the drift of the bound receptors with the retrograde flow velocity, $v_{\rm m}$, as they are attached to F-actin bundle, and the other two terms are the reaction terms as in Eq. \ref{free}. In our model, motivated by the experimental findings, the association and the dissociation rates, $k_{\rm on}(f)$ and $k_{\rm off}(f)$, are considered to be force dependent \cite{Balaban_2001, Tan_PNAS_2003, Rumi_focal_adhesion}. We have taken, $k_{\rm on}(f)= k_{\rm on}^{0} + g(f)\rho_{\rm c}$; where, $k_{\rm on}^{\rm 0}$ is the rate constant, and $g(f)$ is a function of the bond force, $f$; for sake of simplicity it is taken to be linear, $g(f)=\xi f$. 
Thus, the force  increases the binding rate and allows for the formation of new bonds; thus, effectively strengthening the adhesion cluster. Moreover, it has been observed that the force, upto an optimal value, helps strengthen the existing focal adhesion bonds and these force-strengthening molecular bonds are called catch bonds \cite{Thomas_2008, Marshall_2003}. In our model, the dissociation rate of the closed bonds is considered to demonstrate catch behavior as \cite{Pereverzev_catch_2005},$k_{\rm off}=k_{\rm s} e^{f/f_{\rm s}}+k_{\rm c} e^{-f/f_{\rm c}}$., here $k_{\rm 0}$, $k_{\rm s}$, and $k_{\rm c}$ are the rate constants.

Moreover, in our theory, the substrate is considered to be viscoelastic in nature and has been modelled as a spring-dash pot system with spring stiffness $K_{\rm sub}$ and viscosity $\gamma$ as shown in Fig. \ref{model_illustration}. Now, the equation of motion for the substrate is obtained by balancing the total force experienced by all the bonds with the sum of the elastic force $(K_{\rm sub}x_{\rm sub})$ and the viscous drag $(\gamma\dot{x}_{\rm sub})$ of the substrate,

\begin{equation}
\gamma\dot{x}_{\rm sub}+K_{\rm sub}x_{\rm sub}=F_{\rm b}^{\rm total};
\end{equation}
here the total traction force is given as,
$F_{\rm b}^{\rm total}=\int_{0}^{\rm L} f \rho_{\rm c}\left(x\right)dx$.

\section{\label{sec:level1}Dimensionless formulation}
We study the dynamics in dimensionless units. The densities have been scaled as: 
$n_{\rm f}=\frac{\rho_{\rm f}}{\rho_{\rm 0}}$, $n_{\rm c}=\frac{\rho_{\rm c}}{\rho_{\rm 0}}$, where $\rho_{\rm 0}$ is the average density of the receptors and is defined as, $\rho_{\rm 0}=\frac{1}{L} \int_{0}^{\rm L}\left(\rho_{\rm c}+\rho_{\rm f}\right)dx$. The dimensionless time is defined as $\tau=k_{\rm 0}t$, where $k_{\rm on}^{\rm 0}$ is expressed as $\alpha k_{\rm 0}$. Thus, the dimensionless binding and unbinding rates are, $\tilde{k}_{\rm on}=\alpha+\tilde{\xi}\tilde{f}\, n_{\rm c}$ and $\tilde{k}_{\rm off}= \tilde{k}_{\rm off}^{s} \exp\left(\tilde{f}\right) + \tilde{k}_{\rm off}^{\rm c} \exp\left(- \tilde{f}/ \tilde{f}_{\rm c} \right)$, where  $\tilde{\xi}=\frac{\xi f_{\rm s}\rho_{\rm 0}}{k_{\rm 0}}\,$,  $\tilde{k}_{\rm off}^{s}=\frac{k_{\rm off}^{\rm s}}{k_{\rm 0}}$,  $\tilde{k}_{\rm off}^{\rm c}=\frac{k_{\rm off}^{\rm c}}{k_{\rm 0}}$,  $\tilde{f}=\frac{f}{f_{\rm s}}$ and $\tilde{f}_{\rm c}=\frac{f_{\rm c}}{f_{\rm s}}$. The position coordinate is scaled as $X=\frac{x}{x_{\rm 0}}$, where $x_{\rm 0}=\frac{f_{\rm s}}{K_{\rm c}}$. Other dimensionless variables are:$\tilde{D}=\frac{D}{k_{\rm 0}x^{2}_{0}}$, $\tilde{v}_{\rm m}=\frac{v_{\rm m}}{x_{0}k_{\rm 0}}$, $\tilde{\gamma}=\gamma\frac{k_{\rm 0}x_{0}}{f_{\rm s}}$; $\tilde{X}_{\rm sub}=\frac{x_{\rm sub}}{x_{\rm 0}}$; $\tilde{K}_{\rm sub}=\frac{K_{\rm sub}}{K_{c}}$; and $\tilde{F}_{\rm b}^{\rm total}=\frac{F_{\rm b}^{\rm total}}{f_{\rm s}}$.

Thus,  the scaled equations of motion turn out to be,
\begin{eqnarray}
\frac{\partial n_{\rm f}}{\partial\tau} & = & \tilde{D}\frac{\partial^{2}n_{\rm f}}{\partial X^{2}}-\tilde{k}_{\rm on}n_{f}+\tilde{k}_{\rm off}n_{\rm c}\label{freeDimensionless}\\
\frac{\partial n_{\rm c}}{\partial\tau} & = & \tilde{v}_{\rm m}\frac{\partial n_{\rm c}}{\partial X}+\tilde{k}_{\rm on}n_{\rm f}-\tilde{k}_{\rm off}n_{\rm c}\label{closedDimensionless}
\end{eqnarray}
and
\begin{equation}
\tilde{\gamma}\dot{\tilde{X}}_{\rm sub}+\tilde{K}_{\rm sub}\tilde{X}_{\rm sub}  =  \tilde{F}_{\rm b}^{\rm total}.
\label{Subsdimensionless}
\end{equation}

\section{\label{sec:level1}Results}

We have investigated the stick-slip dynamics by  solving  the coupled reaction-diffusion Eqs. \ref{freeDimensionless}-\ref{Subsdimensionless} numerically. The equations are discretized using finite difference method on a grid of size $N$ and then solved by fourth order Runge-Kutta method. The boundary conditions are taken to be such that the total number of free and bound receptors present in the system is conserved. We have studied the dynamics for a wide range of parameter values by varying system size, number of myosin motor, retrograde flow velocity, binding rates, substrate stiffness and viscosity. Here, we present the result for a representative parameter set, where values of the force dependent rate constants are kept at  $\alpha = 2, \tilde \xi = 1, \tilde{k}_{\rm off}^{\rm c} = 120, \tilde{k}_{\rm off}^{\rm s} = 0.25 $ and $\tilde{f}_{\rm c} = 0.5$;  also, the unloaded velocity $\tilde{v}_{\rm 0} = 120$, 
the diffusion constant $\tilde{D}=5$,
 the stall force $F_{m}=2$, number of myosin motors $n_{m}=100$, and the system size $N = 100$. 
The substrate viscosity $\tilde{\gamma}$  and rigidity $\tilde{K}_{\rm sub}$ remain as variable parameters. 
We have also varied the diffusion constant, $\tilde{D}$; the system reaches to the steady state faster with a higher value of diffusion constant, however, the stick-slip dynamics remain the same.  Moreover, we note that even though the model parameters are scaled and dimensionless, nonetheless their values are taken from  experiments, {\it e.g.}, the dissociation rate constants are taken as ${k}_{\rm off}^{\rm c} = 120 \rm \:s^{-1}$ and ${k}_{\rm off}^{\rm s} = 0.25 \rm \:s^{-1}$ \cite{Pereverzev_catch_2005},  whereas, unloaded myosin motor stall force is $F_{\rm m}=2 \rm \:pN $ and unloaded retrograde flow velocity is $v_{0}=120 \rm \:nm\,s^{-1}$ \cite{Chan_Science_2008, Bangasser_biophys_2013}. Also, the variation of substrate elastic stiffness is considered as  $K_{\rm sub}\sim 0.01-100 \: \rm pN\,nm^{-1}$ \cite{Chan_Science_2008, Bangasser_biophys_2013}, and the range of substrate viscosity is $\gamma\sim 0.01-10 \: \rm pN.s\,nm^{-1}$ as observed in experiments \cite{Bennett_PNAS_2018}.

\subsection{\label{sec:level2}Stick-slip dynamics: dependence on substrate stiffness}

\begin{figure}[!t]
\begin{center}
\includegraphics[scale=0.45]{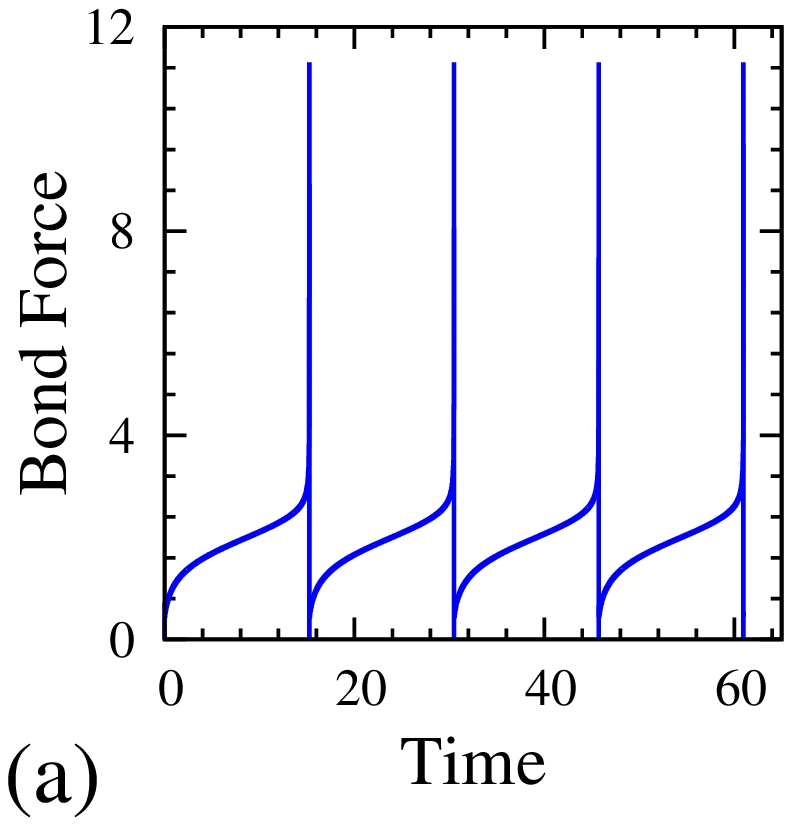}
\includegraphics[scale=0.45]{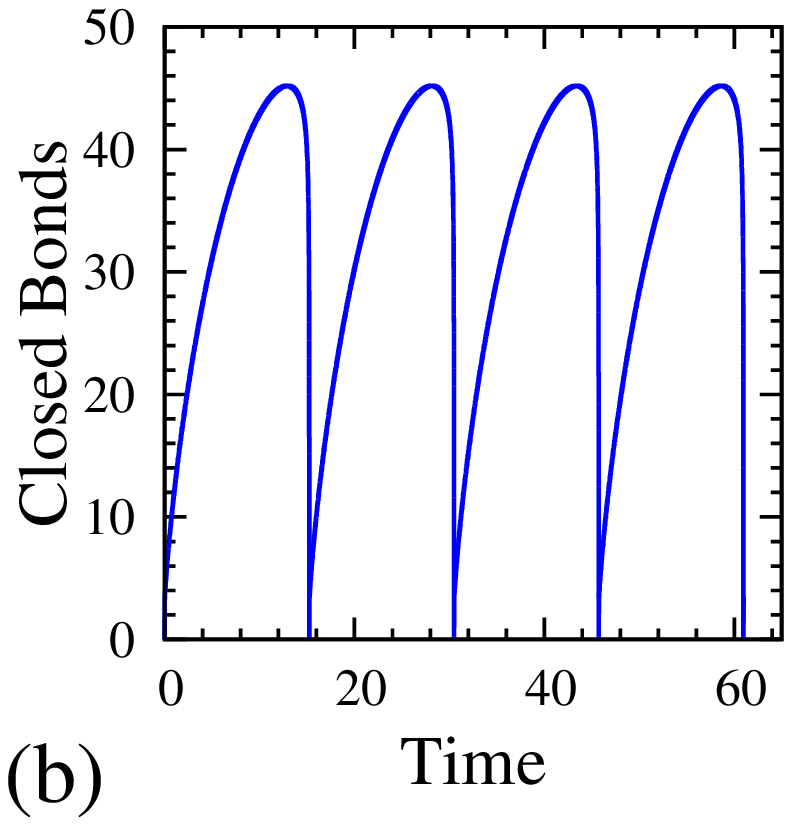}
\includegraphics[scale=0.45]{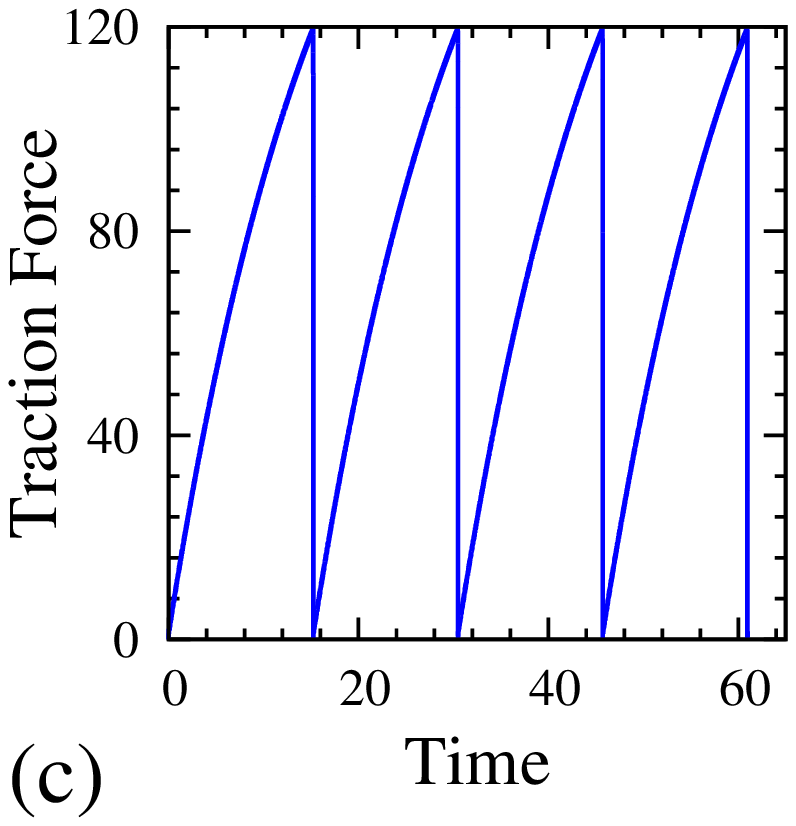}
\includegraphics[scale=0.45]{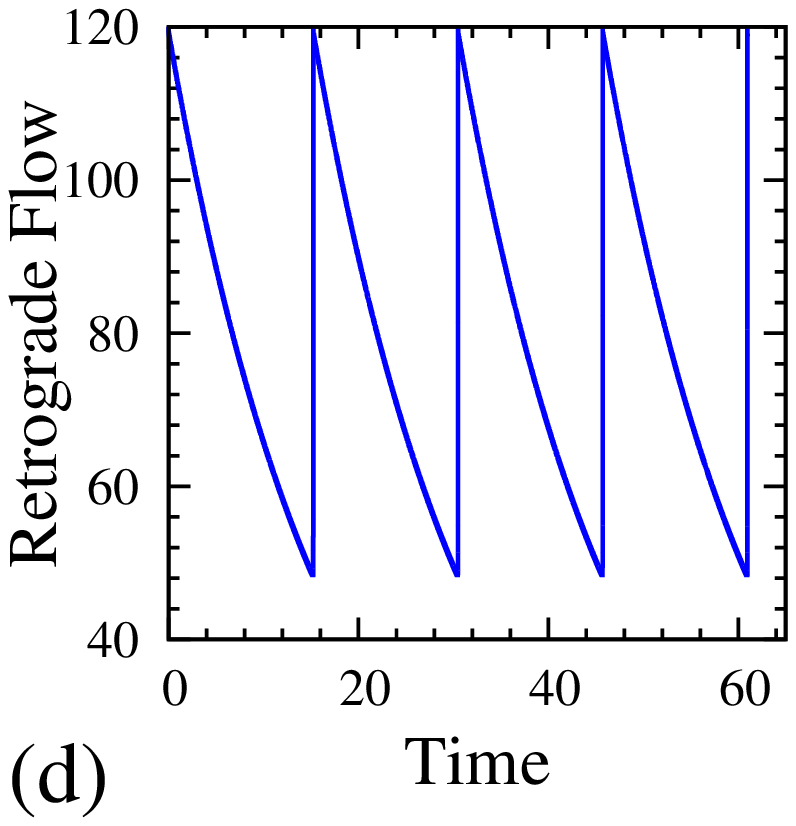}
\caption{Stick-slip motion on soft substrate. (a) Time evolution of single bond force, $\tilde f$. (b) time evolution of the total number of bonds, $N_{\rm c}=\intop_{0}^{\rm L}n_{\rm c}dX$. (c) corresponding evolution of the traction force, $\tilde F_{\rm b}^{\rm total}$, and (d) the retrograde flow velocity, $\tilde v_m$, mirrors the effect of the slowing down of the actin flow with increase in the traction force. (Keeping $\tilde{K}_{\rm sub} = 0.1$ and $\tilde \gamma = 0.01$, all values are dimensionless.)} \label{fig:ksp1_stick_slip}
\end{center}
\end{figure}

Figures \ref{fig:ksp1_stick_slip} a-d show the time evolution of single bond force, total number of bonds, total traction force and corresponding retrograde flow velocity on a soft substrate. Soft substrates are very compliant and deform easily, thus, the build up of force, $\tilde{f}$, on an individual bond is also slow as shown in Fig. \ref{fig:ksp1_stick_slip} a. Now, the increase in bond force, $\tilde{f}$, increases the binding rate, $\tilde{k}_{\rm on}$, of the free receptors and at the same time, decreases the dissociation rate, $\tilde{k}_{\rm off}$, of the bound receptors upto an optimal force value due to catch bond behaviour. As a result, initially a large concentration of receptors are bound to the ligands on the substrate (shown in Fig. \ref{fig:ksp1_stick_slip} b). As the density of the bound receptors increases, they share the  total traction force exerted by the substrate. Thus, the traction force slowly grows with time (Fig. \ref{fig:ksp1_stick_slip} c) as the substrate gets deformed. The growth of traction force in turn slows down the retrograde flow of actin (Fig. \ref{fig:ksp1_stick_slip} d). This gives rise to the \lq stuck' state which allow actin polymerization to advance the leading edge of the cell. But as the force increases even further, the dissociation rate starts to increase. Then, the linearly growing binding rate can no longer keep up and falls below the much faster growing dissociation rate and thus, the adhesion cluster dissociate very quickly and the number of closed bonds decreases. The dissociation of bonds  increases the effective force on the remaining bound receptors/bonds as less number of bonds have to share the currently high value of traction force. This increases the dissociation rate even further and as a result of this feedback cycle, the bound receptors dissociate very quickly. The quick dissociation of bonds means that there is nothing to holding on or anchor to the substrate and thus, the retrograde velocity increases rapidly during what is known as the \lq slip' state as could be seen from Fig. \ref{fig:ksp1_stick_slip} d, and the stick-slip cycle thus repeats.

\begin{figure}[!t]
\begin{center}
\includegraphics[scale=0.45]{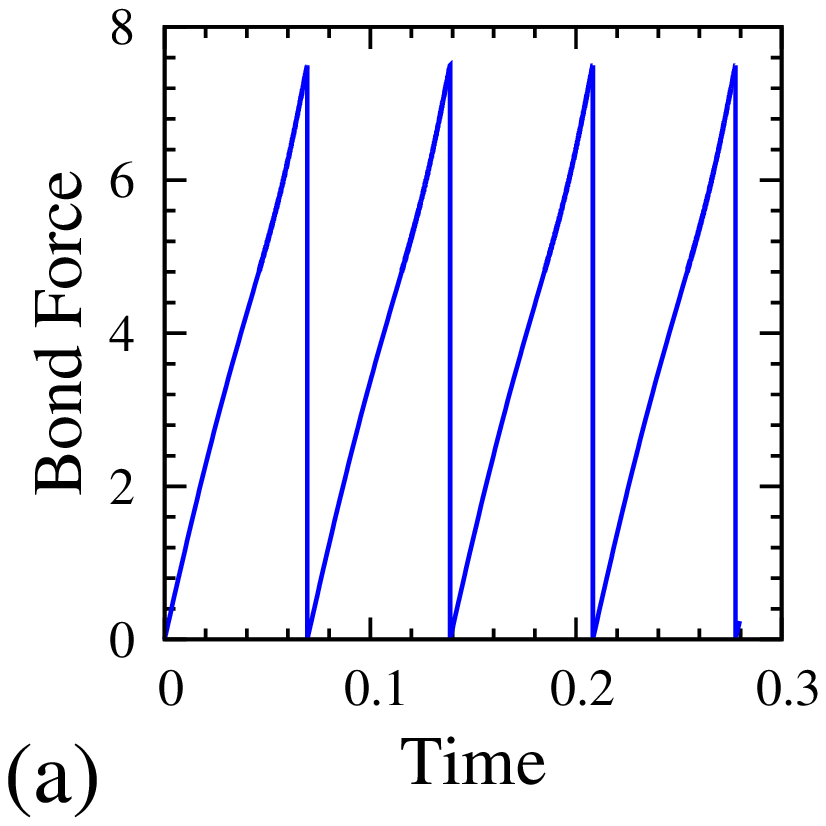}
\includegraphics[scale=0.45]{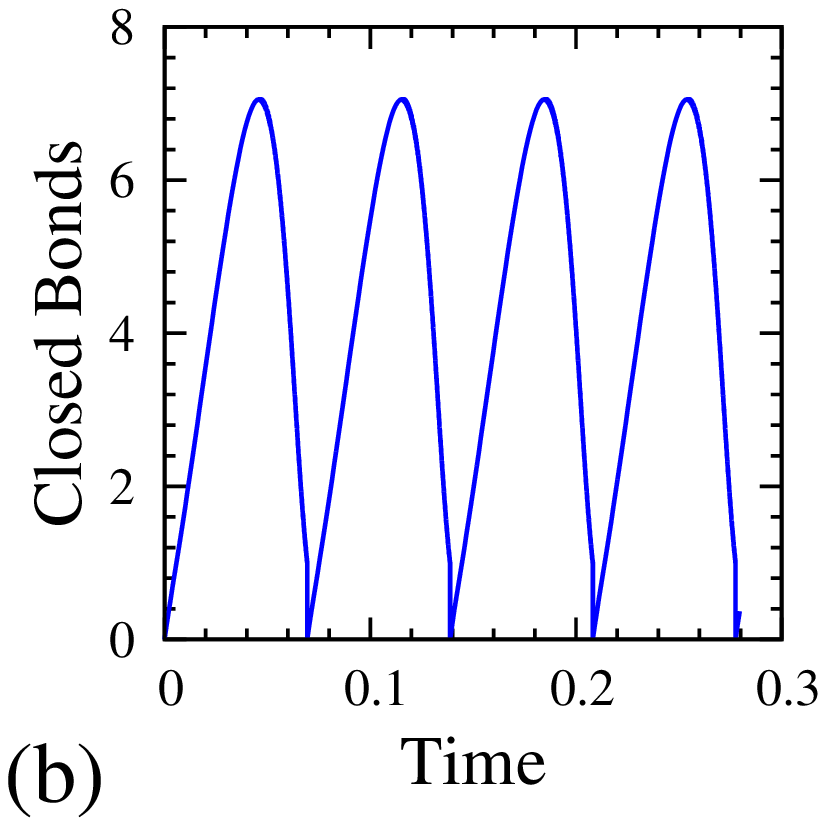}
\includegraphics[scale=0.45]{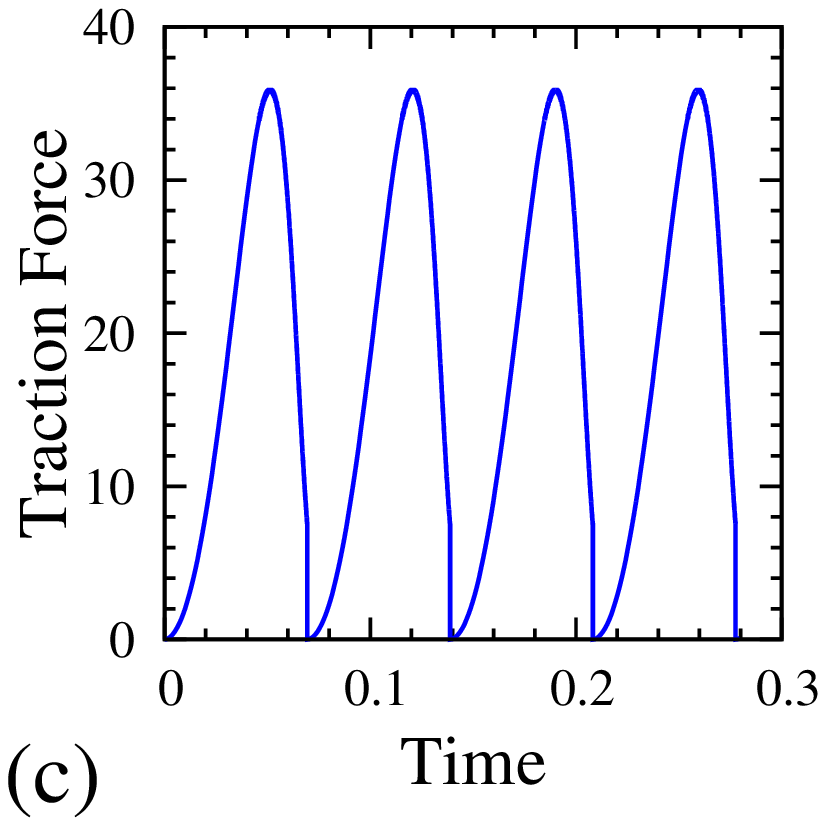}
\includegraphics[scale=0.45]{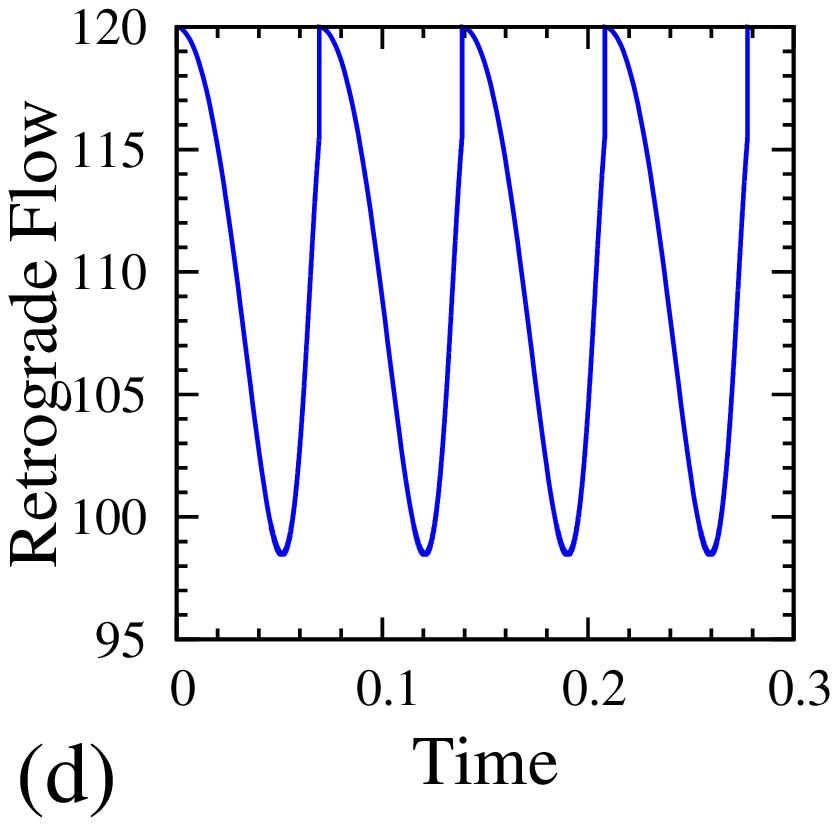}
\caption{Stick-slip motion on stiff substrate. (a) Evolution of single bond force, $\tilde f$, as a function of time. (b) Time evolution of the total number of bonds, $N_{\rm c}=\intop_{0}^{\rm L}n_{\rm c} dX $, (c) corresponding evolution of the traction force, $\tilde F_{\rm b}^{\rm total}$, and (d) the retrograde flow velocity. ( Keeping $\tilde{K}_{\rm sub} = 100$ and $\tilde \gamma =0.01$.)} \label{fig:ks100_stick_slip}
\end{center}
\end{figure}
On the other hand, in case of stiff substrate, as the substrate does not deform easily, force on an individual bond increases very quickly due to actin retrograde velocity as shown in Fig. \ref{fig:ks100_stick_slip} a. However, within this little time, due to lack of sufficient binding time, the formation of bound receptors/bonds happens to be very small as could be seen from Fig. \ref{fig:ks100_stick_slip} b. Now, as these small number of closed bonds are to share the total traction force  exerted by the substrate, this makes the force on a single bond to increase even faster within a short time (Fig. \ref{fig:ks100_stick_slip} a). As a  result, the exponentially varying dissociation rate dominates over linearly increasing binding rate; hence, the adhesions start dissociating even before the substrate has a substantial deformation and the traction force has attained a high value (Fig. \ref{fig:ks100_stick_slip} c). Lower traction force also results in a higher retrograde flow rate (shown in Fig. \ref{fig:ks100_stick_slip} d) as the actin filament slips backward faster and the cell thus is unable to effectively transmit forces to the stiff substrate compared to a soft substrate.

\begin{figure}[!t]
\begin{center}
\includegraphics[scale=0.63]{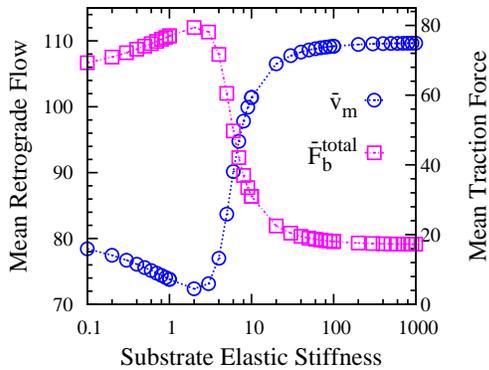}
\caption{Mean value of the retrograde flow velocity $\bar{v}_{\rm m}$ (\protect\markerthree) and thetraction force $\bar{F}_{\rm b}^{\rm total}$ (\protect\markerone) avaraged over the stick-slip cycle as a function of substrate stiffness, $\tilde{K}_{\rm sub}$ (keeping $\tilde\gamma=0.01$ to a constant value). Average traction force is maximum and retrograde flow is minimum for an optimal value of substrate stiffness as observed in experiments \cite{Chan_Science_2008, Bangasser_biophys_2013, Bangasser_natcom_2017}.} \label{fig:Stick_slip_avg}
\end{center}
\end{figure}

We further investigate the stick-slip behaviours as a function  of varying substrate stiffness. As observed in earlier studies \cite{Chan_Science_2008, Bangasser_biophys_2013, Bangasser_natcom_2017}, our simulations show that there exists an optimal substrate stiffness, where the mean value of the total traction force is maximum and the retrograde flow is minimum  as shown in Fig. \ref{fig:Stick_slip_avg}. This could be attributed to the difference in the nature of increase of force of an individual bond depending on the substrate stiffness as evidently seen from Fig \ref{fig:ksp1_stick_slip} a and Fig \ref{fig:ks100_stick_slip} a. On a stiff substrate, since the single bond force increases rapidly, only limited number of bound receptors could form within that short time. Moreover, fast increasing force shortened the life time of the bonds; thus, resulting in lower total traction force and higher retrograde flow. However, on a softer substrate, as the substrate deforms easily, bond force  increases slowly which allows for the formation of more bound receptors. Thus, the higher density of adhesion bonds results in higher value of traction force that resists the actin flow and thus, decreases retrograde velocity. As the substrate is made even more softer, the traction force and consecutively the force on individual bonds grows very slowly, due to the extreme compliance of the substrate. As a result, the system spends a large amount of time in a state of experiencing low traction force and higher retrograde velocity. This reduces the mean value of the traction force for very compliant substrates to some extent and thus increases the mean retrograde velocity.

\begin{figure}[!t]
\begin{center}
\includegraphics[scale=0.65]{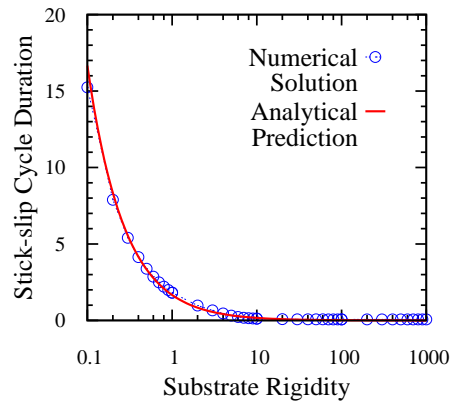}
\caption{Duration of the stick-slip cycle as a function of substrate stiffness, $\tilde{K}_{\rm sub}$ . Numerical results agree well with the analytical prediction of the cycle duration, given by the characteristic time, $\tau_{\rm c}$, obtained from Eq. \ref{time_const} (parameters are kept at same value as of numerical simulation.)}
\label{stickslip_duration}
\end{center}
\end{figure}

Our theory further elucidates  how the variation of substrate stiffness affects the duration of  a stick-slip cycle. We obtain an analytic expression of the time evolution of the total traction force, $\tilde{F}_{\rm b}^{\rm total}(\tau)$ through a rudimentary calculation along with a few approximations (see supporting material). At any instant, the total traction force due to the deformation of the substrate must be balanced by summing over forces of all bound receptors/bonds. Following, the evolution of the total traction force during a stick-slip cycle (starting from a value, $\tilde{F}_{\rm b}^{\rm total}=0$ at time $\tau = 0$) is given by
\begin{equation}
\tilde{F}_{\rm b}^{\rm total} \left(\tau\right) = \tilde{F}_{\rm stall}\left[1-\exp\left(-\frac{\tilde{K}_{\rm c} N_{\rm c} \tilde{v}_{\rm 0} \tilde{K}_{\rm sub}}{\tilde{F}_{\rm stall} \left(\tilde{K}_{\rm sub} + \tilde{K}_{\rm c}N_{\rm c}\right)}\tau\right)\right].
\label{timeevolution_tractionforce}
\end{equation}
Eq. \ref{timeevolution_tractionforce} can be rewritten as,
\begin{equation}
\tilde{F}_{\rm b}^{\rm total} \left(\tau\right) = \tilde{F}_{\rm stall} \left[1-\exp\left(-\frac{\tau}{\tau_{\rm c}}\right)\right];
\end{equation}
where the time constant, $\tau_{\rm c}$, is of the form,
\begin{equation}
\tau_{\rm c} = \frac{\tilde{F}_{\rm stall} \left(\tilde{K}_{\rm sub} + \tilde{K}_{\rm c}N_{\rm c}\right)}{\tilde{K}_{\rm c}N_{\rm c} \tilde{v}_{\rm 0}\tilde{K}_{\rm sub}} = A + \frac{B}{\tilde{K}_{\rm sub}}.
\label{time_const}
\end{equation}
The characteristic time, $\tau_{c}$, is a measure of the growth rate of the total traction force and consequently, slowing down of actin retrograde flow. Therefore, it denotes the time scale corresponding to \lq stuck' state which occupies the majority of the stick-slip cycle duration. As the slip state duration is very small compared to stuck state, the variation of $\tau_{c}$ provides some insights into the duration of the stick-slip cycle on various system parameters, namely, substrate stiffness, bond stiffness, retrograde velocity, myosin activity, and the system size as described by Eq. \ref{time_const}.  Moreover, we also numerically compute the duration of the  stick-slip cycle as a function of substrate stiffness, $\tilde{K}_{\rm sub}$,  as shown in Fig. \ref{stickslip_duration}. Our theoretical prediction (approximated by Eq. \ref{time_const}) matches quite well with the stick-slip cycle duration obtained from simulation results, as seen from the figure Fig. \ref{stickslip_duration}.

\subsection{\label{sec:level2}Effect of the substrate viscosity on the dynamics}

Recent experiments have shown that how cells spread, adhere, migrate or modulate their contractile activity vary with the extracellular matrix depending on whether it is elastic or viscoelastic in nature \cite{Chaudhuri_2015, Lautscham_2014, Bennett_PNAS_2018, Rumi_safran_2010, Rumi_safran_natphy_2007}. Moreover, the substrate stress relaxation is controlled by the viscosity; thus, it  alters the cellular force transmission process and so the overall response of cells. Here, we investigate how the presence of substrate viscosity affects the stick-slip behaviour  of crawling cells.

\begin{figure}[!h]
\begin{center}
\includegraphics[scale=0.48]{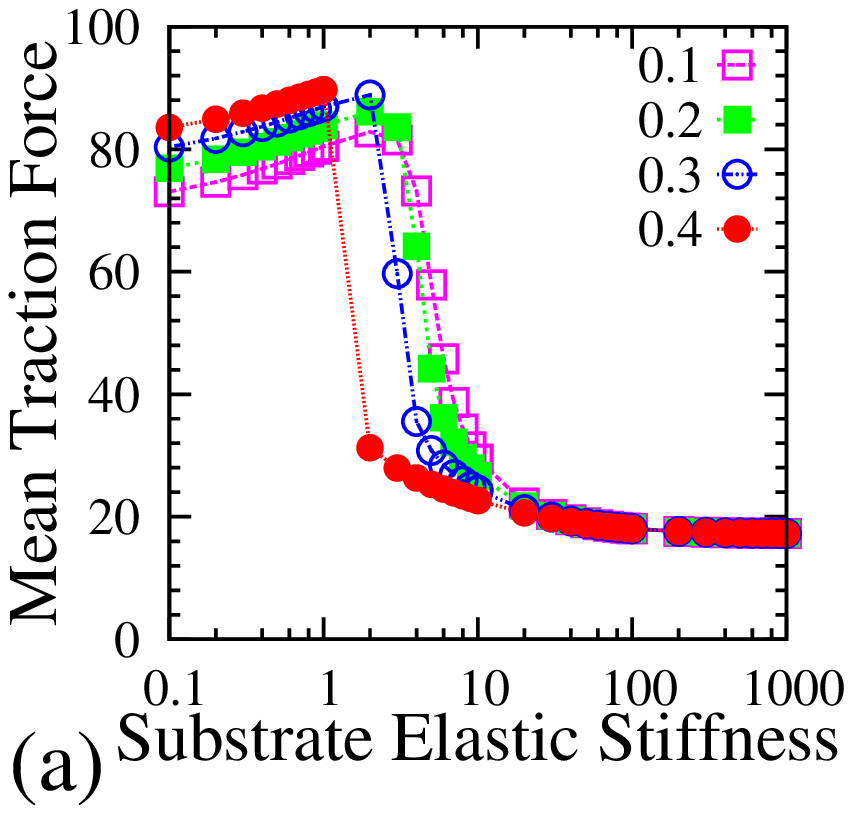}
\includegraphics[scale=0.48]{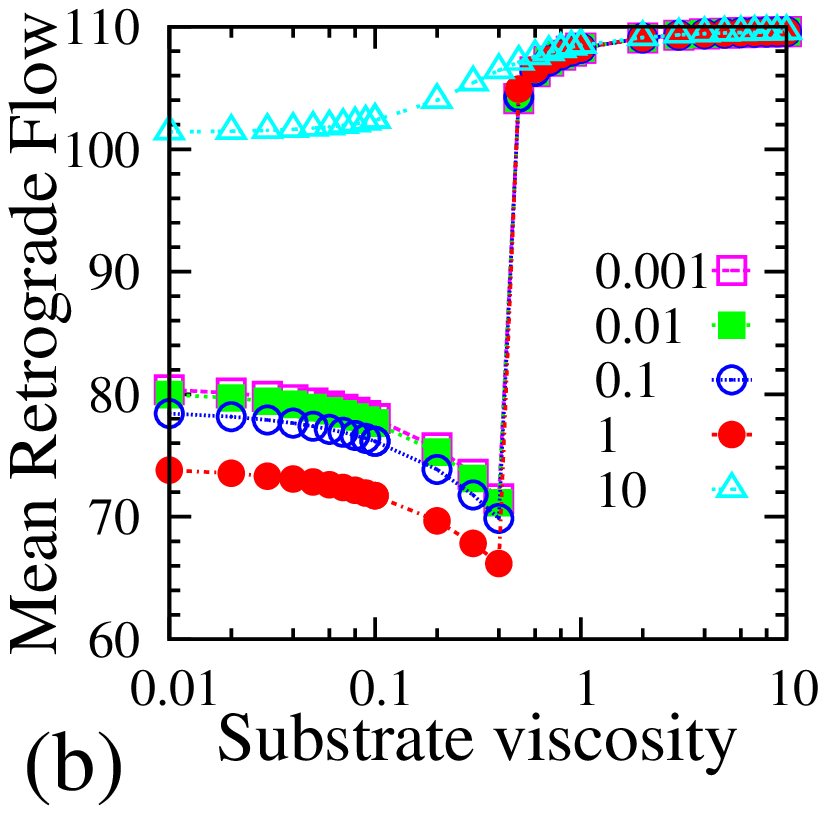}
\caption{(a) Variation of mean traction force, $\tilde F_{\rm b}^{\rm total}$, as a function of substrate stiffness, $\tilde{K}_{\rm sub}$, for different values of substrate viscosity: $\tilde \gamma=0.1$ (\protect\markerone), $0.2$ (\protect\markertwo), $0.3$ (\protect\markerthree), $0.4$ (\protect\markerfour) . (b) Effect of mean retrograde velocity on varying substrate viscosity keeping the elastic stiffness, $\tilde{K}_{\rm sub}$, at a fixed value for different values of elastic stiffness: $\tilde{K}_{\rm sub}=0.001$ (\protect\markerone), $0.01$(\protect\markertwo), $0.1$(\protect\markerthree), $1$(\protect\markerfour), $10$(\protect\markerfive).}
\label{viscosity_plot}
\end{center}
\end{figure}

As shown in Fig. \ref{viscosity_plot} a, increase in viscosity, $\tilde{\gamma}$, increases the effective traction force of the compliant substrate and shifts the optimal substrate stiffness for which the maximum traction occurs to a lower value towards softer, $\tilde{K}_{\rm sub}$. However, for stiff substrates as observed, the presence of viscosity does not cause any noticeable difference in the dynamics because the individual bonds already experience a rapid building of tension due to high elastic stiffness and therefore, destabilize quickly resulting in higher retrograde flow and low stick-slip cycle duration. Thus, at higher stiffness regime, the stick-slip dynamics remain unaltered for elastic and viscoelastic substrate. 

Interestingly, our theory predicts the existence of an optimal substrate viscosity corresponding to maximum traction force and minimum retrograde flow as observed in case of purely elastic substrate. As seen from Fig. \ref{viscosity_plot} b, keeping the substrate elastic stiffness, $\tilde{K}_{\rm sub}$, at a constant value and increasing viscosity, $\tilde{\gamma}$, increases effective traction force, thus, reducing the actin retrograde flow. However, increasing the viscosity further beyond an optimal value,  traction force start decreasing and hence, increases the retrograde  velocity. As the viscous drag resists the deformation of the substrate, the overall stiffness of the substrate increases with increasing viscosity.  At a lower substrate viscosity, since the substrate relaxes faster, force on individual bonds grow slowly allowing longer interaction time with the substrate and formation of more bonds. As the adhesion cluster grows total traction force increases and hence slows down the retrograde flow. On the other hand, at higher substrate viscosity, as the substrate relax slowly, that causes the bond force to rise quickly without providing enough time for the formation of new bonds to share the traction force. As the bond force increases faster, it destabilize the adhesion cluster resulting in lower traction force and higher retrograde velocity as seen from Fig. \ref{viscosity_plot} b.

\subsection{\label{sec:level2}Shifting of optimal stiffness}
\begin{figure}[!t]
\begin{center}
\includegraphics[scale=0.48]{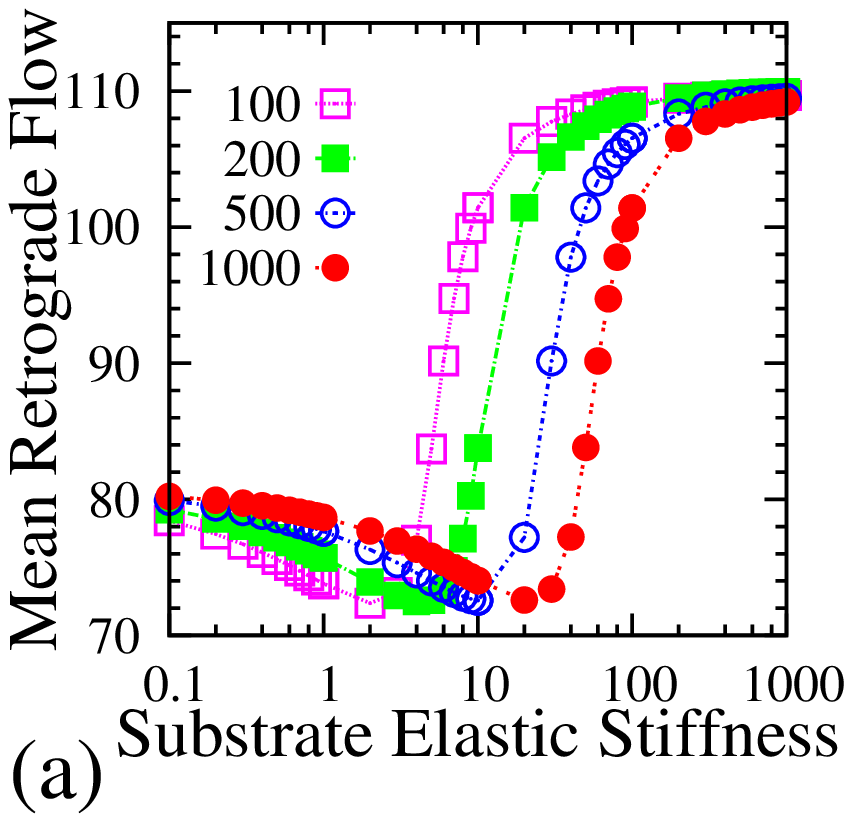}
\includegraphics[scale=0.48]{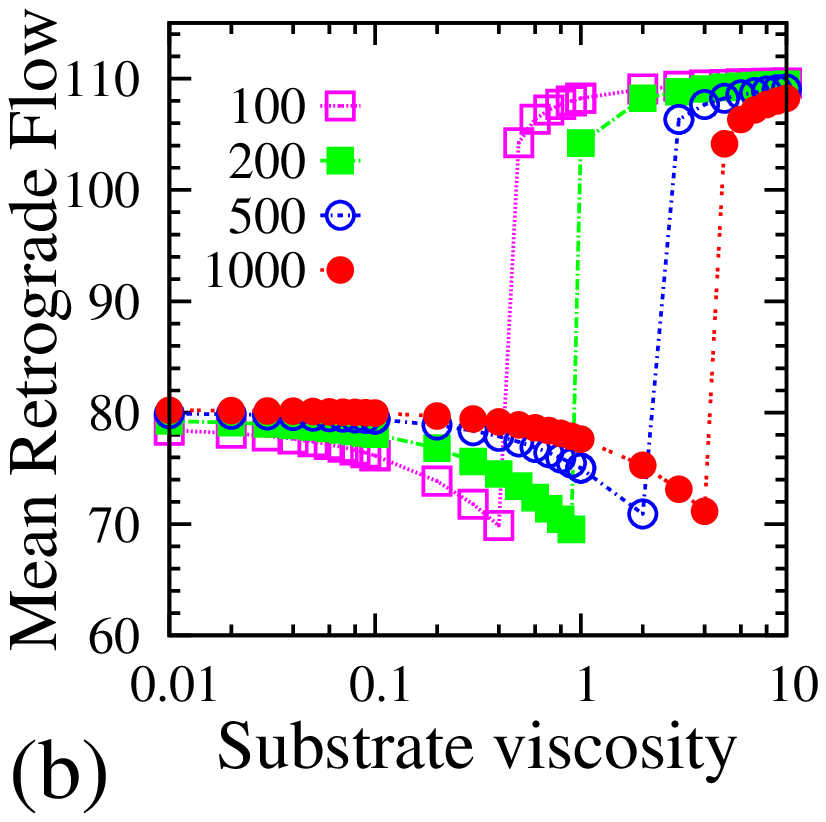}
\caption{Variation in the total number of myosin motors and the size of the adhesion cluster causes shift in (a) the optimal substrate stiffness and also in (b) the optimal viscosity. (Number of motors, $n_{\rm m}$ and total number of receptors, $N_{\rm t}$ have been varied equally.) $n_{\rm m} = N_{\rm t}=100$(\protect\markerone), $200$(\protect\markertwo), $500$(\protect\markerthree), $1000$(\protect\markerfour)}
\label{effect_systemsize_plot}
\end{center}
\end{figure}
We now study how the system parameters such as binding rates, size of the adhesion patch, number of myosin motors affect the optimal substrate viscosity and also compare with that of elastic substrate. It is observed that to exhibit stick-slip behaviour, the pulling force on the F-actin bundle by the myosin motors must be balanced by the total force of the ligand-receptor bonds. However, if  the total number of available receptors  is too less compared to the number of myosin motors, then it results in perpetually \lq slipping' mode, with  F-actin filament moving at near it's unloaded velocity, $v_{0}$. The reverse scenario can also take place where the number of bonds is much higher compared to the number of myosin motors, so that it slows down the retrograde velocity to zero, thus resulting in a permanently \lq stuck' state. However, if the number of myosin motors and the number of bonds, {\it i.e.}, the size of the ligand-receptor adhesive patch are varied appropriately, the stick-slip dynamics is restored.
 
In our simulation, the simultaneous change in number of motors and the size of the adhesive patch is taken into consideration by changing the grid size $N$ and also modifying  $F_{\rm stall} = n_{\rm m}*F_{\rm m}$, where $n_{\rm m}$ denotes the number of myosin motors pulling the F-actin filament.  Fig \ref{effect_systemsize_plot} a and Fig \ref{effect_systemsize_plot} b show the mean retrograde flow for varying system size $N$ and $n_{\rm m}$; where in (a), it is plotted as a function of substrate elastic stiffness, $\tilde K_{\rm sub}$, keeping the viscosity, $\tilde \gamma$, at a constant value and in (b), as a function of substrate viscosity, $\tilde \gamma$, keeping $\tilde K_{\rm sub}$ constant. As seen from Fig. \ref{effect_systemsize_plot} a, with increase in the system size, as more number of receptors/bonds can bind with the ligand on the substrate, the total bond force increases which balances  the  traction force by the substrate; thus, the optimal elastic stiffness for minimum retrograde flow shifts to a higher value. Fig. \ref{effect_systemsize_plot} b shows that increase in the system size also brings  similar shift in the optimal substrate viscosity towards a higher value. This is because, higher viscosity results in slow stress relaxation, thus, increases the effective combined stiffness of the substrate and allows transmission of larger traction force as observed in case of pure elastic stiffness.

We have, further, compared the results for varying binding rates and also studied without the force induced adhesion reinforcement. It is found that the force dependent rate  provides an added flexibility for changing the optimal substrate stiffness or the optimal viscosity to match the experimentally observed values for different cell types. This is achieved because changing the force induced rate results in variation of the total number of closed bonds, thus, changes the total traction force and the retrograde flow velocity and consequently shifts the optimal stiffness.

\section{\label{sec:level1}Discussions}

We have developed a theoretical model to study the \lq stick-slip' motion at the leading edge of a crawling cell. The extracellular matrix has been modelled as a viscoelastic system to better mimic the biological substrates, as opposed to the generally modelled pure elastic substrates. Our continuum model framework comprising of coupled reaction-diffusion equations predicts the time evolution of force on an individual bond during a stick-slip cycle, that could not be captured in existing  stochastic model frameworks. 
Our study reveals that the loading rate of single bond force is distinctively different on soft substrate compared to stiff substrate. It plays a crucial role in determining  the pattern of the stick-slip jumps on varying substrate rigidity. It is also worth noting that our continuum model description reduces the computational  time required for averaging of the dynamical quantities as compared to stochastic models where the averaging needs to be done over a large number of trajectories to extract useful statistical information. 
Also, in our model, motivated by the experimental findings, the bond association and dissociation rates are considered to be force dependent. Experimentally observed force induced reinforcement of adhesion complexes has been incorporated in the binding rate as well as through the \lq catch bond' behaviour of adhesion complexes \cite{Thomas_2008, Marshall_2003, Pereverzev_catch_2005}. 
Moreover, our analysis elucidates the dependence of the duration of the stick-slip cycle on various cellular parameters, for example, how it is affected by myosin activity, retrograde flow, or substrate stiffness. Our theory further suggests that the viscoelasticity of the substrate plays a central role in driving the cell migration process. It reveals the existence  of an \lq optimal' substrate viscosity where the traction force is maximum and the retrograde flow is minimum similar to the variation of elastic substrate stiffness. This indicates the importance of substrate stress relaxation process in cell motility. As in experiments, cell crawling has been found to be most efficient on an optimal substrate stiffness, it could further be tested by altering the viscosity or a combination of both viscosity and elasticity to see how cells respond to viscoelastic tissues and interpreting to the responses to modulate the behaviour in order to fine-tune the biophysical applications such as cancer research \cite{Alberto_natmat_2014}, tissue engineering, regenerative medicine etcetera. \\

\begin{acknowledgements}
The authors acknowledge the financial support from Science and Engineering Research Board (SERB), Grant No. SR/FTP/PS-105/2013, Department of Science and Technology (DST), India.
\end{acknowledgements}

\appendix
\section{Analytical estimation of the duration of a stick-slip cycle:}
Considering an elastic substrate, the traction force due to the deformation of the substrate will be given by $\tilde{F}_{\rm b} ^{\rm total} = \tilde{K}_{\rm sub}\tilde{X}_{\rm sub}$, where $\tilde{X}_{\rm sub}$ denotes the displacement of the substrate and $\tilde{K}_{\rm sub}$ is the substrate stiffness. At any instant, the traction force must be balanced by summing over forces of all ligand-receptor bonds. Now, the total bond force could be calculated as, $\tilde{F}_{\rm b}^{\rm total} = \tilde{K}_{\rm c}N_{\rm c}\left(\tilde{X}_{\rm b} - \tilde{X}_{\rm sub}\right)$, where 
$N_{\rm c}$ is the total number of closed bonds at any instant, $N_{\rm c} = \intop_{0}^{\rm L}n_{\rm c}dX$, the elongation of the bond is given by $\left(\tilde{X}_{\rm b}-\tilde{X}_{\rm sub}\right)$. Here $\tilde{X}_{\rm b}$ is the displacement of one end of the bond attached to the actin filament, thus, $\tilde{X}_{\rm b}=\intop \tilde{v}_{\rm m}d\tau$.
Here, $\tilde{v}_{\rm m}$ is the dimensionless retrograde flow given by $\tilde{v}_{\rm m} = \tilde{v}_{0}(1 - \tilde{F}_{\rm b} ^ {\rm total} /\tilde{F}_{\rm stall})$.

Now, the expression for the traction force at any time $\tau$ can be rewritten as,
\begin{equation}
\tilde{F}_{\rm b}^{\rm total} = \tilde{K}_{\rm c}N_{\rm c}\left(\tilde{X}_{\rm b} - \frac{\tilde{F}_{\rm b}^{\rm total}}{\tilde{K}_{\rm sub}}\right)
\label{dimensionless_1}
\end{equation}
Differentiating Eq. \ref{dimensionless_1} w.r.t. the dimensionless time $\tau$ and using the relation,
$ \frac{d\tilde X_{\rm b}}{d\tau} = \tilde{v}_{\rm m}$, we obtain,
\begin{equation}
 \frac{d\tilde{F}_{\rm b}^{\rm total}} {d\tau}\left(1 + \frac{\tilde{K}_{\rm c}N_{\rm c}}{\tilde{K}_{\rm sub}}\right) = \tilde{K}_{\rm c}N_{\rm c}\tilde{v}_{0}\left(1 - \frac{\tilde{F}_{\rm b} ^{\rm total}}{\tilde{F}_{\rm stall}}\right)\label{dimensionless_2}
 \end{equation} 
Thus, the evolution of the total traction force during a stick-slip cycle starting from a value $0$ at time $\tau = 0$ can be given by
the solution of Eq. \ref{dimensionless_2},
\begin{equation}
\tilde{F}_{\rm b}^{\rm total}\left(\tau\right) = \tilde{F}_{\rm stall} \left[1 - \exp\left(-\frac{\tilde{K}_{\rm c}N_{\rm c} \tilde{v}_{0}\tilde{K}_{\rm sub}}{\tilde{F}_{\rm stall}\left(\tilde{K}_{\rm sub} + \tilde{K}_{\rm c}N_{\rm c}\right)}\tau\right)\right].
\label{dimensionless_3}
\end{equation}

\end{document}